\documentclass[aps,pra,groupedaddress,superscriptaddress,nofootinbib]{revtex4}

\usepackage{graphicx,amsmath,amssymb,amsbsy,subfigure,hyperref,bbm,times,txfonts,float}
\usepackage{subfigure,hyperref,bbm,times}
\usepackage[T1]{fontenc}
\usepackage{braket}
\usepackage{epsfig} 
\usepackage{color}
\usepackage{amsmath}
\usepackage{xcolor}%
\usepackage{graphicx}
\usepackage{dcolumn}
\usepackage{bm}

\providecommand{\openone}{\leavevmode\hbox{\small1\kern-3.8pt\normalsize1}}

\usepackage{soul}

\usepackage{xcolor}
\usepackage[normalem]{ulem}

\hypersetup{
   colorlinks=true,
   linkcolor=blue,
}
\graphicspath{{figure/}}

\begin{document}

\title{Kerr‑Induced Control of Synchronization and Quantum State Recovery in a Driven van der Pol Oscillator}
 \author{Amir Hossein Houshmand Almani}
\affiliation{Department of Physics, University of Guilan, P. O. Box 41335-1914, Rasht, Iran}
 \author{Ali Mortezapour}
 \affiliation{Department of Physics, University of Guilan, P. O. Box 41335-1914, Rasht, Iran}
 \author{Alireza Nourmandipour}
  \email{anourmandip@sirjantech.ac.ir}
 \affiliation{Department of Physics, Sirjan University of Technology, 7813733385 Sirjan, Iran}

\begin{abstract}
	We investigate how Kerr nonlinearity modifies quantum synchronization in a squeezed quantum van der Pol oscillator. We show that the Kerr interaction produces an amplitude-dependent frequency shift that drives a saddle-node bifurcation, transforming the classical phase-space structure from bistable to monostable dynamics. In the quantum regime, this transition manifests as systematic frequency pulling and spectral broadening, while the steady-state Wigner function reveals a continuous correspondence between the quantum state and the semiclassical attractor despite finite quantum fluctuations. By constructing global synchronization phase diagrams in the squeezing--Kerr parameter space, we uncover a remarkably linear dependence of the critical squeezing strength required to maintain phase locking on the Kerr nonlinearity. We further demonstrate that the synchronization boundary does not coincide with the crossover between super- and sub-Poissonian photon statistics, showing that synchronization and photon-number statistics characterize distinct aspects of the quantum steady state. These results provide quantitative design principles for controlling quantum synchronization through Kerr nonlinearity, with potential relevance to trapped-ion, superconducting-circuit, and optomechanical platforms.
\end{abstract}

\date{\today }

\maketitle

\section{Introduction}

Synchronization is a ubiquitous manifestation of nonlinear dynamics, describing the adjustment of rhythms between interacting or externally driven oscillatory systems. It occurs across a broad range of physical, biological, and technological settings, including Josephson junctions, laser arrays, mechanical resonators, neuronal networks, and chemical oscillators \cite{1,2,3,4,5,6}. Beyond its fundamental importance, synchronization underpins applications in precision metrology, coherent signal generation, frequency stabilization, and information processing \cite{7,8,9,10,11}.

The extension of synchronization to the quantum regime has attracted considerable attention because quantum fluctuations fundamentally modify the mechanisms underlying phase locking \cite{Houshmand2025Annalen,12,13,14,15}. Unlike classical oscillators, quantum systems are subject to unavoidable fluctuations arising from dissipation, measurement back-action, and the uncertainty principle, making robust synchronization a nontrivial problem. Understanding how nonlinear interactions compete with quantum fluctuations has therefore become an important theme in quantum optics, optomechanics, trapped ions, superconducting circuits, and driven-dissipative quantum systems \cite{HoushmandAlmani2025,16,17,18,19,20,HoushmandAlmani2026}.

A paradigmatic platform for investigating quantum synchronization is the van der Pol (vdP) oscillator. The classical vdP oscillator supports self-sustained oscillations through the balance between negative linear damping and nonlinear dissipation. Its quantum counterpart is naturally described by a Lindblad master equation in which incoherent single-photon pumping and two-photon loss realize the corresponding gain and nonlinear damping processes \cite{12,13}. As a minimal model of a self-oscillating quantum system, the quantum vdP oscillator has become a standard framework for studying synchronization in the presence of quantum fluctuations. Previous studies have shown that synchronization can persist deep in the quantum regime, although quantum noise progressively degrades the quality of phase locking and broadens the associated spectral response \cite{20,24,25}.

Several strategies have therefore been explored to enhance quantum synchronization. Among them, parametric squeezing has emerged as a particularly powerful resource \cite{26,27}. A two-photon squeezing drive can produce strong frequency entrainment and, in contrast to conventional harmonic driving, can yield a synchronization range that is less sensitive to the dissipative parameters \cite{26}. The resulting improvement in phase locking and spectral coherence makes squeezing a promising resource for synchronization in trapped-ion, optomechanical, superconducting, and nonlinear photonic platforms.

Realistic quantum platforms, however, rarely operate in the absence of additional nonlinearities. A particularly ubiquitous example is the optical Kerr effect, which originates from the third-order nonlinear susceptibility $\chi^{(3)}$ and produces an intensity-dependent refractive index and, consequently, an amplitude-dependent resonance frequency \cite{28,29}. In the rotating frame, the Kerr interaction appears as a nonlinear Hamiltonian proportional to $\mathcal{K}\hat n^2$, leading to an amplitude-dependent frequency shift that modifies the resonance condition of the oscillator. Because synchronization relies on phase locking to an external reference, Kerr-induced frequency pulling is expected to compete directly with squeezing-induced phase stabilization. Despite the ubiquity of Kerr nonlinearities in quantum and photonic platforms, the nonlinear competition between strong Kerr interactions and squeezing-driven synchronization in the deep quantum regime remains largely unexplored.

In this work, we develop a comprehensive semiclassical and quantum description of synchronization in a Kerr-driven squeezed quantum van der Pol oscillator. We show that the Kerr interaction generates an amplitude-dependent effective detuning that reorganizes the synchronization dynamics and drives a saddle-node bifurcation, transforming the classical phase-space structure from bistable to monostable behavior. In the quantum regime, the same mechanism manifests itself through systematic frequency pulling, spectral broadening, and continuous deformation of the steady-state Wigner function. By constructing global quantum phase diagrams in the $(\eta,\mathcal{K})$ parameter space, we identify quantitative synchronization boundaries and uncover a remarkably simple linear scaling of the critical squeezing strength with the Kerr nonlinearity. We further show that the synchronization boundary does not coincide with the crossover between super- and sub-Poissonian photon statistics, demonstrating that photon-number statistics do not uniquely determine the onset of synchronization.

The remainder of this paper is organized as follows. Section~\ref{model} introduces the model and derives the corresponding semiclassical equations of motion. In Sec.~\ref{sec:semiclassical}, we analyze the semiclassical fixed points, their stability, and the Kerr-induced saddle-node bifurcation. Section~\ref{sec:phase_boundaries} then investigates the corresponding quantum dynamics through the steady-state Wigner function, emission spectrum, and quantitative synchronization phase diagrams, including the dependence of the synchronization boundaries on detuning and coherent driving. Finally, Sec.~\ref{sec:photon_statistics} examines the photon statistics across the synchronization transition through the Mandel $Q_M$ parameter. Our main conclusions are summarized in Sec.~\ref{sec:conclusion}.

\section{The Model}
\label{model}

We consider a quantum van der Pol oscillator whose synchronization dynamics are simultaneously influenced by two competing nonlinear mechanisms: parametric squeezing and Kerr nonlinearity. The starting point for our investigation is the classical vdP equation, which captures the self-sustained limit-cycle behavior resulting from the interplay between negative linear damping and positive nonlinear damping \cite{30,31}:
\begin{equation}
	\ddot{x} - \mu(1 - x^2)\dot{x} + \omega_0^2 x = F \cos(\omega_d t),
	\label{eq:classical_vdp}
\end{equation}
where \(\omega_0\) is the natural frequency, \(\omega_d\) is the drive frequency, \(F\) is the driving amplitude, and \(\mu\) governs the nonlinearity. The term \(-\mu(1-x^2)\dot{x}\) provides negative damping at small amplitudes and positive damping at large amplitudes, giving rise to self‑sustained limit‑cycle oscillations. When driven by a periodic force, this system can synchronize its rhythm to the external stimulus, a phenomenon central to many classical and quantum technologies.

To explore the quantum regime, we quantize the oscillator mode using the bosonic annihilation operator \(\hat{b}\). In a frame rotating at the drive frequency \(\omega_d\), the dynamics of the system's density matrix \(\rho\) are governed by a Lindblad master equation \cite{12,13}:
\begin{equation}
	\dot{\rho} = -i[\hat{H}_{\text{tot}}, \rho] + \gamma_1 \mathcal{D}[\hat{b}^\dagger]\rho + \gamma_2 \mathcal{D}[\hat{b}^2]\rho .
	\label{eq:master}
\end{equation}
Here, \(\mathcal{D}[\hat{O}]\rho = \hat{O}\rho\hat{O}^\dagger - \frac{1}{2}\{\hat{O}^\dagger\hat{O}, \rho\}\) represents the standard dissipator. The rates \(\gamma_1\) and \(\gamma_2\) characterize the linear pumping (gain) and two-photon nonlinear damping, respectively. The interplay of these two dissipative terms stabilizes the quantum limit cycle analogous to the classical vdP oscillator.

The coherent dynamics are encapsulated in the total Hamiltonian \(\hat{H}_{\text{tot}}\), which we define as:
\begin{equation}
	\hat{H}_{\text{tot}} = \Delta \hat{b}^\dagger \hat{b} + iF(\hat{b} - \hat{b}^\dagger) + i\eta(\hat{b}^2 e^{-i\theta} - \hat{b}^{\dagger 2} e^{i\theta}) + \mathcal{K}(\hat{n}^2 - \hat{n}).
	\label{eq:Hamiltonian}
\end{equation}
The first term accounts for the detuning \(\Delta = \omega_0 - \omega_d\) between the oscillator and the external drive. The second term represents a coherent harmonic drive of strength \(F\). The third term, parameterized by amplitude \(\eta\) and phase \(\theta\), describes a two-photon squeezing drive. The final term represents the Kerr nonlinearity, which generates an intensity-dependent frequency shift and consequently shears the oscillator phase space. The coupling constant \(\mathcal{K}\) (proportional to \(\chi^{(3)}\)) induces an intensity-dependent frequency shift.

To gain analytical insight into the competition between squeezing and Kerr effects, we derive semiclassical equations of motion. In the high-excitation limit (\(\gamma_1 \gg \gamma_2\)), the photon number is macroscopically large, allowing us to neglect quantum fluctuations. We adopt the semiclassical ansatz \(\langle \hat{b} \rangle = R e^{i\varphi}\) and factorize higher-order moments as \(\langle \hat{b}^\dagger\hat{b}\,\hat{b} \rangle \approx R^3 e^{i\varphi}\) and \(\langle \hat{n}\hat{b} \rangle \approx R^3 e^{i\varphi}\). Substituting these into the expectation value evolution derived from Eq.~\eqref{eq:master}, and separating real and imaginary parts, yields the coupled equations for the amplitude \(R\) and phase \(\varphi\):
\begin{align}
	\dot{R} &= \frac{\gamma_1}{2} R - \gamma_2 R^3 - F \cos\varphi - 2\eta R \cos(2\varphi - \theta), \label{eq:Rdot} \\
	\dot{\varphi} &= -\Delta - 2\mathcal{K} R^2 + \frac{F}{R} \sin\varphi + 2\eta \sin(2\varphi - \theta). \label{eq:phidot}
\end{align}

With the sign convention adopted in Eq.~\eqref{eq:Hamiltonian}, a positive Kerr coefficient corresponds to a self-focusing nonlinearity, producing an amplitude-dependent blue shift of the oscillator frequency. Consequently, the effective detuning becomes \(\Delta_{\text{eff}} = \Delta + 2\mathcal{K} R^2\). As the oscillation amplitude increases, the Kerr interaction progressively shifts the oscillator away from resonance, thereby opposing the phase-locking action of the squeezing drive.

\section{Semiclassical Fixed Points and Bifurcation Analysis}
\label{sec:semiclassical}

In this section, we investigate the stationary solutions of the semiclassical equations of motion and analyze how the Kerr nonlinearity modifies the phase-space structure of the driven dissipative oscillator. The stationary states are determined by solving the nonlinear algebraic equations
\begin{align}
	\dot{R} &= 0,
	\\
	\dot{\phi} &= 0,
\end{align}
which define the equilibrium points in the $(R,\phi)$ phase space.

\begin{figure}[t]
	\centering
	\includegraphics[width=\textwidth]{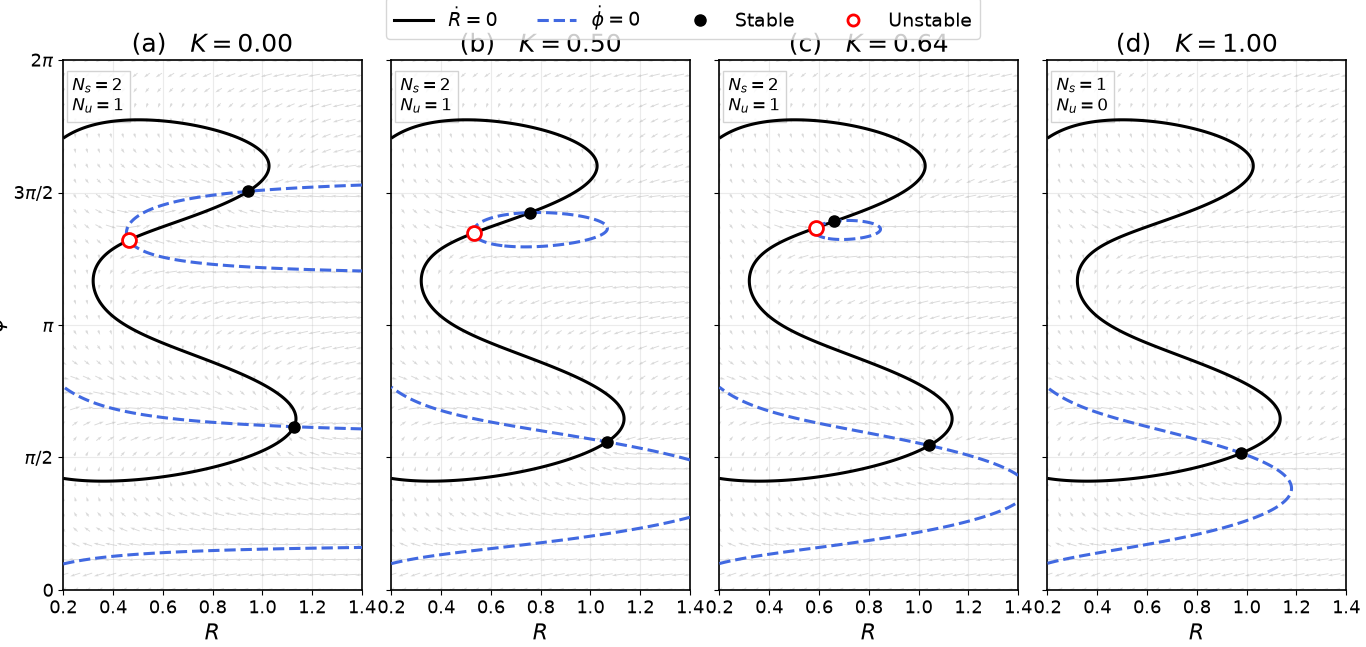}
	\caption{
		(Color online) Phase portraits of the classical Kerr--van der Pol oscillator in the $(R,\phi)$ phase space for different Kerr nonlinearities:
		(a) $\mathcal{K}=0$,
		(b) $\mathcal{K}=0.50$,
		(c) $\mathcal{K}\simeq\mathcal{K}_{\mathrm{SN}}$, and
		(d) $\mathcal{K}=1.0$.
		The gray arrows denote the normalized vector field.
		The solid black curves and blue dashed curves represent the amplitude and phase nullclines, defined by $\dot{R}=0$ and $\dot{\phi}=0$, respectively.
		Filled black circles indicate stable fixed points, whereas open red circles denote unstable fixed points.		
	}
	\label{Fig1}
\end{figure}

The local stability of each equilibrium point is determined from the eigenvalues of the Jacobian matrix,
\begin{equation}
	J=
	\begin{pmatrix}
		\partial_R \dot{R} &
		\partial_{\phi} \dot{R}
		\\
		\partial_R \dot{\phi} &
		\partial_{\phi} \dot{\phi}
	\end{pmatrix},
	\label{eq:Jacobian}
\end{equation}
evaluated at the corresponding stationary solution. An equilibrium is classified as stable when both eigenvalues have negative real parts, whereas the existence of at least one eigenvalue with a positive real part indicates an unstable stationary state.

Figure~\ref{Fig1} presents the phase portraits of the semiclassical dynamics for representative values of the Kerr nonlinearity $\mathcal{K}$. The gray arrows represent the normalized vector field, while the solid black and dashed blue curves correspond to the amplitude and phase nullclines, respectively, defined by $\dot{R}=0$ and $\dot{\phi}=0$. Their intersections identify the stationary solutions. Filled black circles denote stable fixed points, whereas open red circles indicate unstable ones.

 Throughout this section we set the squeezing drive amplitude and the coherent drive amplitude as \(\eta/\gamma_1 = 1.5\) and \(F/\gamma_1 = 1\), with a detuning \(\Delta/\gamma_1 = 1\). These values are chosen so that, in the absence of the Kerr effect (\(\mathcal{K}=0\)), the system exhibits three fixed points: two stable nodes and an unstable saddle, characteristic of a bistable regime arising from a saddle‑node bifurcation. The effect of \(\mathcal{K}\) is then isolated and monitored.

For weak Kerr nonlinearity, the system possesses three equilibrium points, consisting of two stable attractors separated by one unstable saddle. As the Kerr strength increases, the phase nullcline is continuously displaced, whereas the amplitude nullcline remains nearly unchanged. This behavior reflects the fact that the Kerr nonlinearity modifies only the phase equation, leaving the amplitude equation unaffected.

As a consequence, one stable equilibrium and the unstable saddle progressively approach each other until they coalesce at a critical Kerr strength, denoted by $\mathcal{K}_{\mathrm{SN}}$. At this point, a saddle-node bifurcation occurs, leading to the annihilation of the two equilibria. Beyond the critical value ($\mathcal{K}>\mathcal{K}_{\mathrm{SN}}$), only a single stable stationary state survives, and the phase-space topology changes qualitatively from a bistable to a monostable regime.

The semiclassical bifurcation revealed in Fig.~\ref{Fig1} constitutes the classical counterpart of the quantum synchronization transition analyzed in the remainder of this work. As we demonstrate below, the disappearance of one classical attractor is accompanied by a substantial restructuring of the quantum steady state, ultimately leading to enhanced quantum synchronization.

The physical origin of this transition can be understood from the Kerr-induced nonlinear phase shift. In the phase-locked regime, the oscillator must satisfy a delicate balance between the intrinsic detuning and the phase pulling generated by the external drive. The Kerr nonlinearity contributes an amplitude-dependent frequency shift through the term $-2\mathcal{K}R^{2}$, which can be interpreted as an effective detuning, i.e., $\Delta_{\mathrm{eff}} = \Delta + 2\mathcal{K}R^{2}$.
%
%
Unlike the bare detuning, the effective detuning depends on the oscillation amplitude itself, introducing a nonlinear feedback between the amplitude and phase dynamics. As the oscillation amplitude changes, the resonance condition is modified self-consistently, causing the phase-locking condition to evolve with the state of the oscillator.

For weak Kerr nonlinearity, this self-consistent balance admits three stationary solutions, comprising two stable attractors separated by an unstable saddle. As $\mathcal{K}$ increases, the effective detuning becomes larger, continuously shifting the phase nullcline while leaving the amplitude nullcline nearly unchanged. Consequently, one stable equilibrium and the unstable saddle approach each other until they coalesce at the critical Kerr strength, $\mathcal{K}_{\mathrm{SN}}$. Beyond this point, the two equilibria annihilate through a saddle-node bifurcation, leaving a single stable synchronized state.

\begin{figure*}[t]
	\centering
	\includegraphics[width=\textwidth]{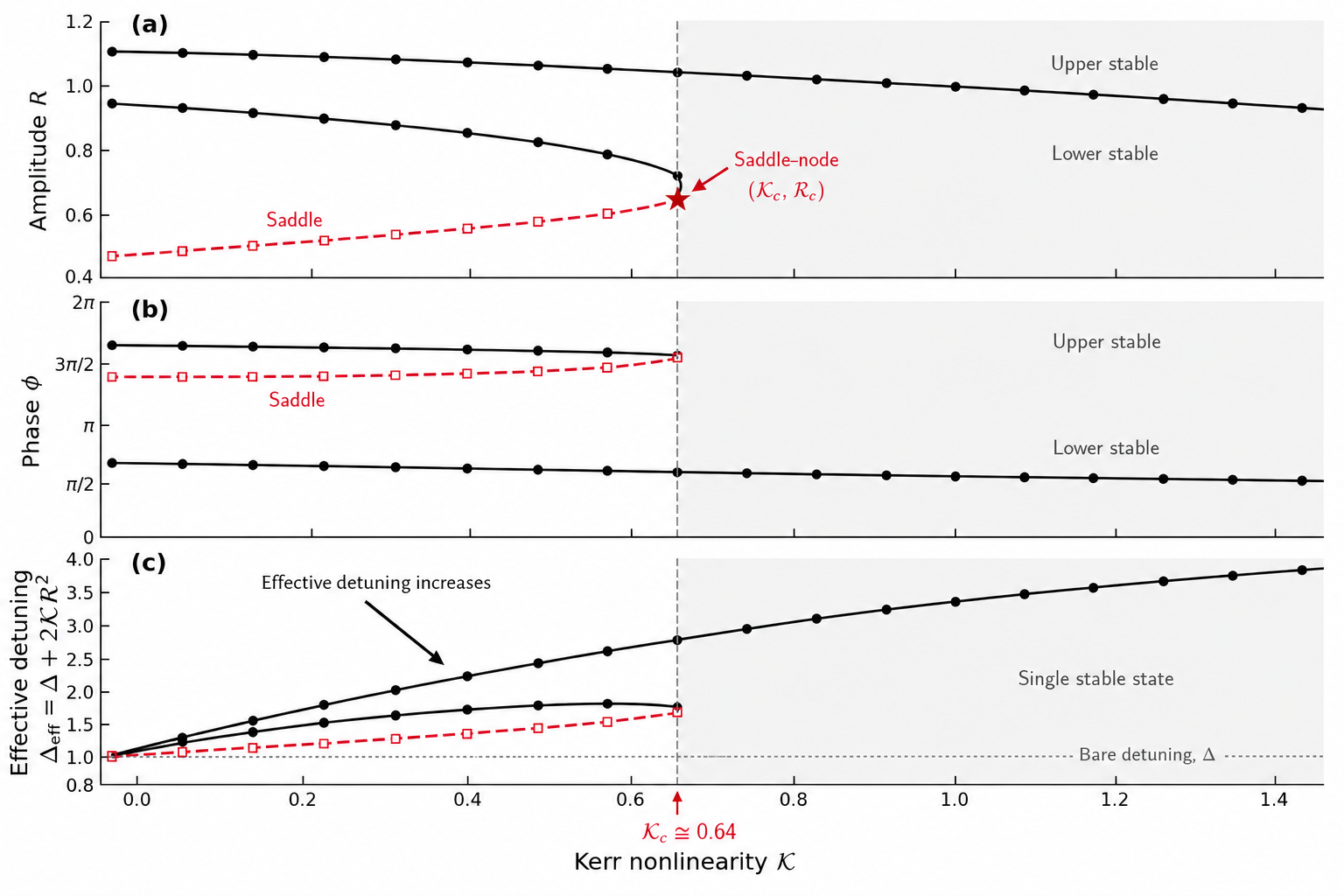}
	\caption{
		(Color online)
		Semiclassical bifurcation induced by the Kerr nonlinearity.
		(a) Amplitudes of the stationary solutions as functions of the Kerr strength $\mathcal{K}$.
		Solid black curves denote stable equilibria, while the red dashed curve represents the unstable saddle branch.
		The star marks the saddle-node bifurcation occurring at the critical Kerr strength $(\mathcal{K}_{\mathrm{c}},R_{\mathrm{c}})$.
		(b) Corresponding phase coordinates of the stationary solutions.
		(c) Effective detuning,
		$\Delta_{\mathrm{eff}}=\Delta+2\mathcal{K}R^{2}$,
		evaluated at each equilibrium point.
		The horizontal dotted line indicates the bare detuning $\Delta$, while the shaded region corresponds to the parameter regime where only a single stable equilibrium remains.
	}
	\label{Fig2}
\end{figure*}

To quantify this transition, Fig.~\ref{Fig2} follows the evolution of the stationary solutions as the Kerr nonlinearity is continuously increased. Panel~(a) shows the amplitudes of all equilibrium branches as functions of the Kerr strength. Two stable branches (solid black) coexist with an unstable saddle branch (red dashed) for small $\mathcal{K}$, confirming the bistable nature of the semiclassical dynamics. As $\mathcal{K}$ increases, the lower stable branch continuously approaches the saddle branch until both collide at the critical Kerr strength, $\mathcal{K}_{\mathrm{c}}\simeq0.64$, where a saddle-node bifurcation occurs. Beyond this point only the upper stable branch survives, indicating that the system has entered a monostable regime.

Panel~(b) displays the corresponding phase coordinates of the equilibrium solutions. The phase of the upper stable branch remains close to $\phi\simeq\pi/2$, whereas the lower stable branch stays near $\phi\simeq3\pi/2$. The saddle follows the latter branch and merges with it at $\mathcal{K}_{\mathrm{c}}$, consistent with the bifurcation observed in the amplitude dynamics.

The physical mechanism responsible for this transition becomes evident in panel~(c), which depicts the effective detuning, $\Delta_{\mathrm{eff}} = \Delta + 2\mathcal{K}R^{2}$ evaluated at each stationary solution. Since the Kerr contribution is proportional to both the nonlinear strength and the stationary oscillation amplitude, increasing $\mathcal{K}$ continuously shifts the resonance condition experienced by each equilibrium. The upper branch, possessing the largest amplitude, exhibits the strongest Kerr-induced frequency shift, while the lower stable branch and the saddle undergo considerably weaker shifts. Consequently, the phase nullcline is progressively displaced relative to the amplitude nullcline, causing the lower stable equilibrium and the saddle to approach each other until they annihilate through the saddle-node bifurcation.

The horizontal dotted line in Fig.~\ref{Fig2}(c) denotes the bare detuning $\Delta$, highlighting the increasing contribution of the Kerr-induced nonlinear frequency shift. Therefore, the disappearance of bistability is not simply a numerical consequence of increasing $\mathcal{K}$, but rather originates from the self-consistent growth of the effective detuning, which continuously modifies the phase-locking condition of the driven oscillator. This interpretation provides a direct physical explanation for the geometric evolution observed in Fig.~\ref{Fig1}. As shown in the following sections, this classical bifurcation leaves clear fingerprints in the quantum steady state, including the deformation of the Wigner function, the restructuring of the emission spectrum, and the emergence of quantitative synchronization phase boundaries.

\section{Quantum Phase Diagram and Synchronization Boundaries}
\label{sec:phase_boundaries}

The semiclassical analysis presented in Sec.~\ref{sec:semiclassical} establishes the existence of synchronized steady states and identifies the Kerr-induced saddle-node bifurcation governing their stability. Although this description captures the underlying nonlinear dynamics, it neglects quantum fluctuations arising from dissipation and finite excitation numbers. Since synchronization in the present system ultimately emerges from a quantum steady state governed by the Lindblad master equation, it is essential to determine how the semiclassical predictions manifest in the fully quantum regime.

To address this question, we employ three complementary diagnostics. First, the steady-state Wigner quasiprobability distribution provides a phase-space representation of the quantum state and allows a direct comparison with the classical attractor. Second, the emission spectrum obtained from the quantum regression theorem characterizes the dynamical coherence of the synchronized oscillator through its frequency, linewidth, and spectral intensity. Finally, by systematically exploring the two-dimensional parameter space of Kerr nonlinearity and squeezing strength, we construct quantitative synchronization phase diagrams and extract the corresponding critical phase boundaries. Together, these analyses establish a consistent correspondence between the semiclassical bifurcation picture and the observable quantum dynamics.

The steady-state Wigner function is obtained from the stationary density operator $\hat{\rho}_{\rm ss}$ satisfying the Lindblad master equation and is defined as \cite{32}

\begin{equation}
	W(\alpha)=
	\frac{2}{\pi}
	\mathrm{Tr}
	\left[
	\hat D(\alpha)
	\hat\rho_{\rm ss}
	\hat D^\dagger(\alpha)
	(-1)^{\hat a^\dagger\hat a}
	\right],
	\label{eq:wigner}
\end{equation}

where $\hat D(\alpha)$ denotes the displacement operator. Throughout this work, the steady-state density matrix is computed numerically using the QuTiP framework \cite{33,34,35}, from which the corresponding Wigner distributions are evaluated.

To compare the quantum steady state with its semiclassical counterpart, we examine three characteristic quantities in phase space. The first is the semiclassical fixed point,

\begin{equation}
	\alpha_s
	=
	R_s e^{i\phi_s},
	\label{eq:classical_alpha}
\end{equation}

whose amplitude and phase are determined from the steady-state solutions of the classical equations of motion. The second quantity is the quantum mean field,

\begin{equation}
	\langle\hat a\rangle
	=
	\mathrm{Tr}
	\left(
	\hat a\hat\rho_{\rm ss}
	\right),
	\label{eq:meanfield}
\end{equation}

which represents the centroid of the quantum state in phase space. Finally, we determine the location at which the Wigner function reaches its maximum,

\begin{equation}
	\alpha_{\rm peak}
	=
	\underset{\alpha}{\mathrm{arg\,max}}
	\,
	W(\alpha),
	\label{eq:wignerpeak}
\end{equation}

corresponding to the most probable phase-space configuration of the oscillator.

\begin{figure*}[t]
	\centering
	\includegraphics[width=0.95\textwidth]{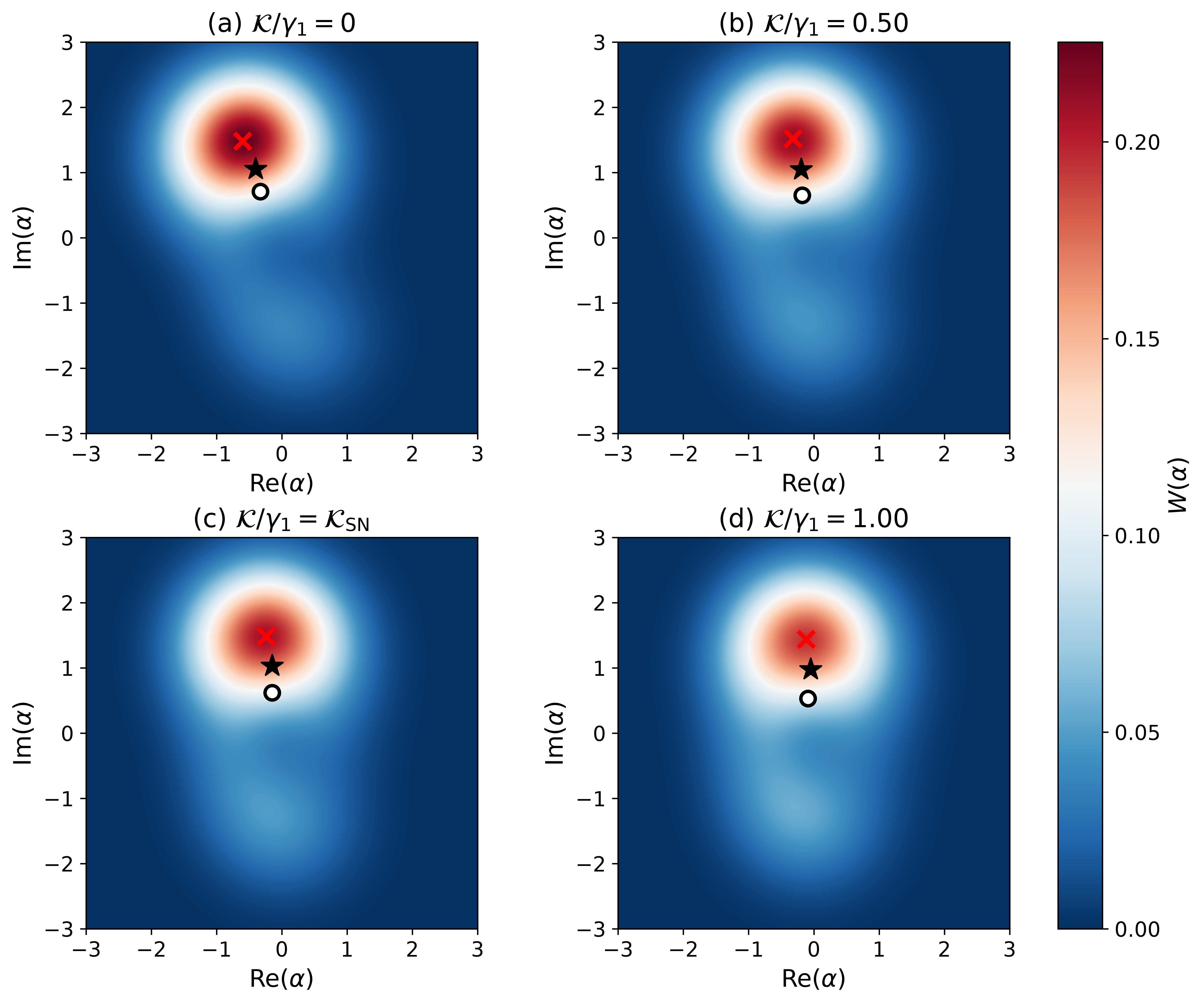}
	\caption{
		 Classical--quantum correspondence in phase space for the Kerr-driven quantum van der Pol oscillator. The steady-state Wigner quasiprobability distributions are shown for four representative values of the normalized Kerr nonlinearity: (a) $\mathcal{K}/\gamma_1=0$, (b) $\mathcal{K}/\gamma_1=0.50$, (c) $\mathcal{K}/\gamma_1=\mathcal{K}_{\rm SN}$, and (d) $\mathcal{K}/\gamma_1=1.00$. The black star denotes the semiclassical stable fixed point,
		$\alpha_s=R_s e^{i\phi_s}$,
		obtained from the classical equations of motion. The white circle indicates the quantum mean field
		$\langle\hat a\rangle=\mathrm{Tr}(\hat a\hat\rho_{\rm ss})$,
		while the red cross marks the position
		$\alpha_{\rm peak}$,
		where the Wigner function reaches its maximum value.
	}
	\label{Fig3}
\end{figure*}

Figure~\ref{Fig3} illustrates the evolution of the steady-state Wigner function for four representative Kerr strengths spanning the synchronized regime. The semiclassical fixed point, the quantum mean field, and the maximum of the Wigner distribution are indicated by the black star, white circle, and red cross, respectively. For all investigated parameters, the Wigner function remains positive and strongly localized, indicating that the steady state retains a predominantly semiclassical character despite the presence of nonlinear dissipation and Kerr interactions. 
	
	As the Kerr strength increases, however, the Wigner distribution gradually deforms and its centroid shifts through phase space. Consequently, finite separations develop between the semiclassical fixed point, the quantum expectation value, and the maximum of the Wigner distribution. Such deviations are expected because, unlike an ideal coherent state whose Gaussian Wigner function possesses coincident centroid and maximum, the anharmonic Kerr interaction generates asymmetric distortions of the quantum state. The resulting offsets therefore quantify the influence of quantum fluctuations and nonlinear phase-space deformation rather than indicating a qualitative breakdown of the semiclassical description.
	
	Nevertheless, the evolution of all three quantities remains remarkably consistent. Their systematic displacement follows the same trajectory predicted by the semiclassical fixed-point analysis, demonstrating that the quantum steady state continuously inherits the classical saddle-node structure while exhibiting finite quantum corrections.
	
	Although the Wigner function provides a detailed static characterization of the quantum steady state, it does not directly reveal the dynamical coherence of the synchronized oscillator. To characterize the temporal properties of the steady state, we therefore examine its emission spectrum. The emission spectrum is particularly attractive because it is directly accessible experimentally through heterodyne or homodyne detection, thereby providing an observable signature of synchronization. The power spectrum is calculated from the steady-state two-time correlation function using the quantum regression theorem implemented in QuTiP. Unlike the phase-space representation, the spectrum directly probes the characteristic oscillation frequency together with the coherence time of the emitted field.
	
\begin{figure*}[t]
	\centering
	\includegraphics[width=0.55\textwidth]{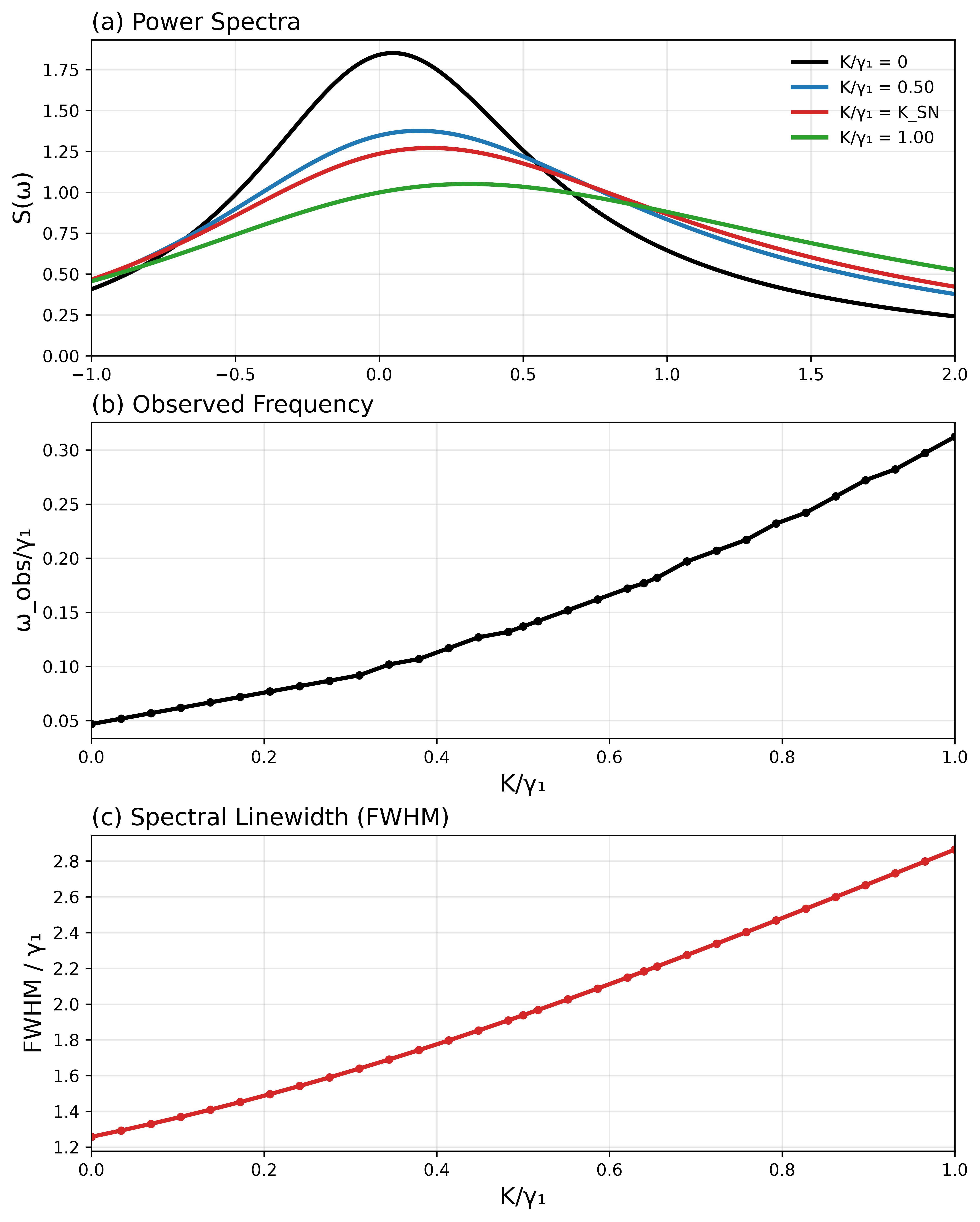}
	\caption{
Spectral signatures of the Kerr-induced synchronization shift.
(a) Power spectra \(S(\omega)\) for increasing Kerr strengths \(\mathcal{K}/\gamma_1 = 0.00, 0.50, 0.64,\) and \(1.00\).
(b) Observed oscillation frequency \(\omega_{\mathrm{obs}}\), extracted from the maximum of the power spectrum versus Kerr strength \(\mathcal{K}/\gamma_1\).
(c) Full width at half maximum (FWHM) of the power spectra.
	}
	\label{Fig4}
\end{figure*}

Figure~\ref{Fig4}(a) displays the emission spectra for the same representative Kerr strengths considered in Fig.~\ref{Fig3}. As the Kerr nonlinearity increases, the spectral maximum shifts continuously toward higher frequencies while simultaneously becoming broader and less pronounced.

The extracted peak frequencies are summarized in Fig.~\ref{Fig4}(b). The observed frequency increases monotonically with Kerr strength, providing direct evidence of Kerr-induced frequency pulling. This behavior is consistent with the effective detuning, $\Delta_{\rm eff}=\Delta+	2\mathcal{K}R^2$  predicted by the semiclassical theory, where the nonlinear Kerr interaction continuously shifts the oscillator away from the external driving frequency.

The corresponding spectral linewidths, quantified by the full width at half maximum (FWHM), are shown in Fig.~\ref{Fig4}(c). The linewidth broadens substantially as the Kerr strength increases, indicating progressively shorter coherence times of the emitted field. Simultaneously, the spectral peak height decreases because the spectral weight becomes distributed over a broader frequency interval. These observations demonstrate that increasing Kerr nonlinearity not only modifies the oscillation frequency but also reduces the spectral coherence of the synchronized state.

Taken together, the spectral evolution complements the phase-space analysis of Fig.~\ref{Fig3}. Whereas the Wigner function visualizes the gradual deformation of the quantum steady state, the emission spectrum reveals its dynamical consequences through frequency pulling and linewidth broadening.

\begin{figure*}[t]
	\centering
	\includegraphics[width=0.75\textwidth]{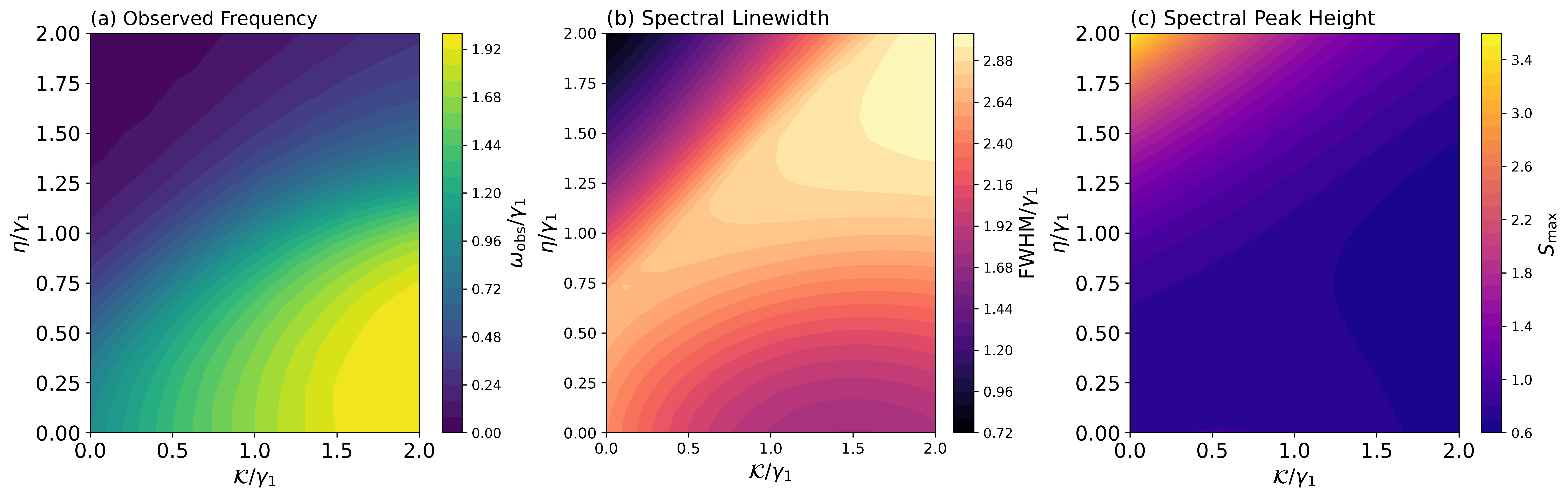}
	\caption{Quantum phase diagrams of the Kerr--squeezing competition in the deep quantum regime. The horizontal and vertical axes denote $\mathcal{K}/\gamma_1$ and $\eta/\gamma_1$, respectively. (a) Observed oscillation frequency $\omega_{\mathrm{obs}}$, (b) spectral linewidth (FWHM), and (c) spectral peak height.
	}
	\label{Fig5}
\end{figure*}

While Fig.~\ref{Fig4} characterizes synchronization at several representative operating points, it does not reveal how synchronization evolves throughout parameter space. We therefore extend the analysis to the full two-dimensional parameter space spanned by the squeezing strength $\eta$ and the Kerr nonlinearity $\mathcal{K}$. Rather than examining isolated operating points, this approach reveals how the competition between these two nonlinear processes organizes the global synchronization landscape of the quantum oscillator.

Figure~\ref{Fig5}(a) presents the observed synchronization frequency $\omega_{\rm obs}$ extracted from the maximum of the steady-state emission spectrum. A well-defined diagonal crossover separates two qualitatively different dynamical regimes. In the upper-left region, where the squeezing drive dominates over the Kerr-induced frequency shift, the oscillator remains phase locked to the external drive and the observed frequency stays close to zero in the rotating frame. As the Kerr nonlinearity increases, however, the effective detuning $\Delta_{\rm eff}=\Delta+2\mathcal{K}R^2$
gradually exceeds the locking capability of the squeezing drive, causing the emission peak to move continuously toward higher frequencies. The smooth nature of this crossover is consistent with the gradual quantum remnant of the saddle-node bifurcation identified in the semiclassical analysis.

The corresponding spectral coherence is illustrated in Fig.~\ref{Fig5}(b), which maps the full width at half maximum (FWHM) of the emission spectrum. Throughout the synchronized region the linewidth remains narrow, indicating long-lived phase coherence. In contrast, once the Kerr-induced detuning dominates, the linewidth increases rapidly, signaling enhanced phase diffusion and the progressive loss of coherent synchronization. The close correspondence between the linewidth broadening and the frequency transition demonstrates that the synchronization crossover is accompanied by a substantial reduction of spectral coherence.

Figure~\ref{Fig5}(c) displays the peak height of the emission spectrum. The strongest and sharpest spectral peaks are found inside the synchronized region, whereas the peak intensity decreases significantly as the oscillator enters the Kerr-dominated regime. This reduction follows naturally from the simultaneous broadening of the spectrum: as the spectral weight spreads over a wider frequency interval, the maximum spectral intensity necessarily decreases. Consequently, the three panels together provide mutually consistent signatures of the same physical transition. The synchronized phase is characterized by a nearly locked emission frequency, narrow linewidth, and high spectral intensity, whereas the unsynchronized regime exhibits a shifted emission frequency, substantial linewidth broadening, and suppressed peak amplitude.

\begin{figure*}[t]
	\centering
	\includegraphics[width=0.75\textwidth]{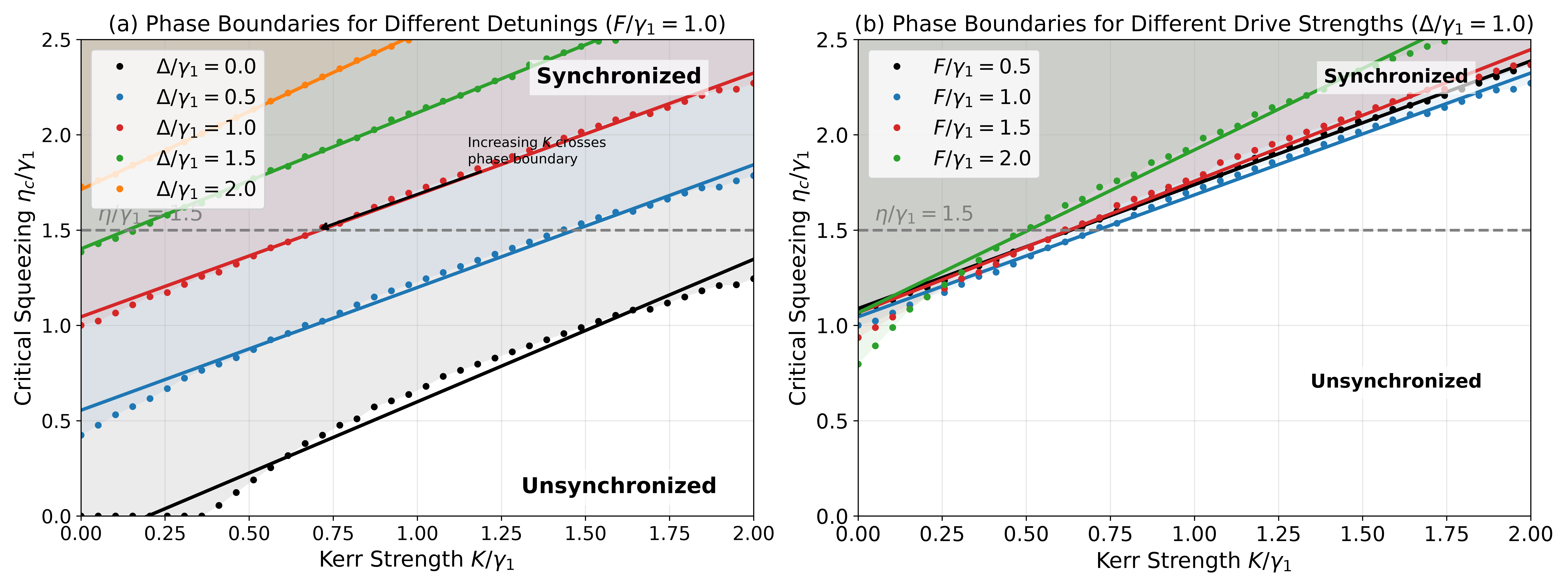}
	\caption{
		Quantitative phase boundaries of quantum synchronization extracted from the two-dimensional phase diagrams of Fig.~\ref{Fig5}. The critical squeezing strength $\eta_c$ is defined by the criterion $\omega_{\mathrm{obs}}=\omega_{\mathrm{th}}=0.2\gamma_1$. Symbols denote the numerically extracted boundaries, and solid lines represent linear least-squares fits. 
		(a) Synchronization boundaries for different detunings $\Delta$.
		(b) Synchronization boundaries for different coherent drive strengths $F$. 
		Shaded regions indicate the synchronized regime.
	}
	\label{Fig6}
\end{figure*}

Although Fig.~\ref{Fig5} clearly reveals the crossover between the synchronized and Kerr-dominated regimes, the location of this transition remains qualitative. To obtain a predictive characterization of the synchronization threshold, we extract the critical squeezing strength $\eta_c(\mathcal{K})$. The resulting phase boundaries are presented in Fig.~\ref{Fig6}, where the symbols denote the numerically extracted thresholds and the solid lines represent linear least-squares fits. Across the investigated parameter range, the boundaries are well described by the approximately linear relation
\begin{equation}
	\eta_{\mathrm{c}}(\mathcal{K})
	=
	\eta_{\mathrm{c},0}
	+
	m\mathcal{K},
	\label{eq:critical_squeezing}
\end{equation}
where $\eta_{\mathrm{c},0}$ denotes the critical squeezing strength in the absence of Kerr nonlinearity and $m$ quantifies the sensitivity of the synchronization threshold to the Kerr strength.

Figure~\ref{Fig6}(a) demonstrates that increasing the cavity detuning shifts the synchronization boundary almost rigidly toward larger squeezing strengths. Physically, a larger bare detuning increases the effective frequency mismatch between the oscillator and the external reference, requiring stronger parametric squeezing to restore phase locking. Remarkably, despite this displacement, the phase boundary remains nearly linear over the entire parameter range investigated.

A different behavior emerges when the coherent drive amplitude is varied, as shown in Fig.~\ref{Fig6}(b). In this case, the intercept of the synchronization boundary changes only weakly, whereas its slope increases systematically with the driving strength. Stronger coherent driving populates the oscillator with more photons, thereby enhancing the Kerr-induced nonlinear frequency shift. Consequently, the squeezing required to compensate this nonlinear detuning grows more rapidly with increasing Kerr strength.

The nearly linear scaling observed in both panels indicates that the competition between Kerr-induced frequency pulling and parametric squeezing is governed by a simple quantitative balance throughout the deep quantum regime. This result extends the semiclassical picture developed in Sec.~\ref{sec:semiclassical} by providing an experimentally accessible synchronization criterion that directly relates the Kerr nonlinearity, squeezing strength, cavity detuning, and driving amplitude. More importantly, it establishes a practical calibration curve that can be used to predict the minimum squeezing required to maintain quantum synchronization for arbitrary operating conditions.

\section{Photon Statistics Across the Synchronization Transition}
\label{sec:photon_statistics}

The results presented in the previous section established that synchronization in the Kerr-driven quantum van der Pol oscillator is governed by the competition between Kerr-induced frequency pulling and squeezing-induced phase locking. The synchronization boundary was identified from the spectral response of the oscillator and shown to obey a remarkably simple linear scaling over a broad parameter regime. An important remaining question, however, concerns the role of quantum fluctuations in this transition. In particular, one may ask whether the onset of synchronization is accompanied by a corresponding change in the photon statistics of the steady state. If synchronization were directly associated with the formation of a coherent optical field, one might expect the synchronization boundary to coincide with a transition between classical and nonclassical photon-number statistics.

To address this question, we characterize the steady state through the Mandel parameter \cite{36},
\begin{equation}
	Q_M
	=
	\frac{\langle \hat n^2\rangle
		-
		\langle \hat n\rangle^2
		-
		\langle \hat n\rangle}
	{\langle \hat n\rangle},
	\label{eq:mandel}
\end{equation}
where $\hat n=\hat b^\dagger\hat b$ is the photon-number operator and the expectation values are evaluated with respect to the steady-state density operator $\hat{\rho}_{\rm ss}$. The Mandel parameter measures deviations from Poissonian photon statistics: $Q_M<0$ indicates sub-Poissonian statistics associated with reduced photon-number fluctuations and nonclassical light, whereas $Q_M>0$ corresponds to super-Poissonian statistics characterized by enhanced fluctuations.

\begin{figure*}[t]
	\centering
	\includegraphics[width=\textwidth]{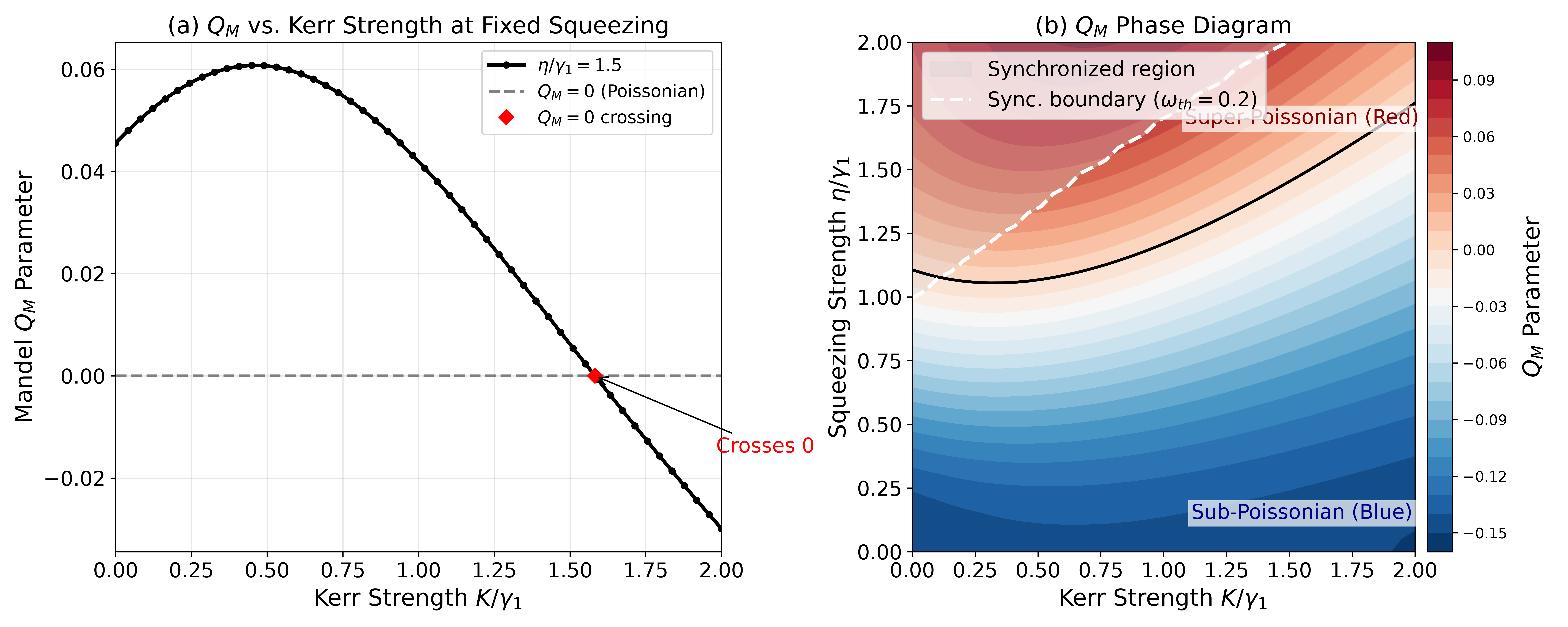}
\caption{
	(Color online) Mandel $Q_M$ parameter across the Kerr--squeezing parameter space.
	(a) $Q_M$ as a function of the Kerr strength $\mathcal{K}/\gamma_1$ for fixed squeezing strength $\eta/\gamma_1=1.5$; the red diamond marks the $Q_M=0$ crossing.
	(b) Mandel parameter over the $(\mathcal{K},\eta)$ parameter space. The white dashed line denotes the synchronization boundary defined by $\omega_{\mathrm{obs}}=0.2\gamma_1$, while the solid black contour indicates $Q_M=0$. The shaded region represents the synchronized regime.
}
	\label{Fig7}
\end{figure*}

Figure~\ref{Fig7}(a) shows the evolution of the Mandel parameter as the Kerr nonlinearity increases for a representative squeezing strength of $\eta/\gamma_1=1.5$. The dependence is smooth over the entire parameter range. Starting from a weakly super-Poissonian state at small Kerr strengths, the Mandel parameter initially increases slightly before decreasing monotonically, eventually becoming negative for sufficiently strong Kerr nonlinearities. Increasing the Kerr interaction therefore progressively suppresses the photon-number fluctuations and drives the steady state into a sub-Poissonian regime. Importantly, this crossover occurs continuously, without any abrupt feature that could be associated with the synchronization boundary identified from the spectral observables.

To obtain a global picture, Fig.~\ref{Fig7}(b) presents the Mandel parameter throughout the $(\mathcal{K},\eta)$ parameter space. The synchronization boundary extracted from the spectral analysis of Sec.~\ref{sec:phase_boundaries} is superimposed as a dashed white curve, while the contour corresponding to $Q_M=0$ is shown as a solid black line, separating the super-Poissonian and sub-Poissonian regions. Several important observations emerge. First, the Mandel parameter varies smoothly throughout the parameter space, exhibiting no abrupt change across the synchronization boundary. Second, the synchronization boundary cuts across regions possessing both positive and negative Mandel parameters. Most significantly, the dashed white synchronization boundary and the solid black $Q_M=0$ contour follow entirely different trajectories. Consequently, synchronized steady states may exhibit either super-Poissonian or sub-Poissonian photon statistics depending on the operating point, while unsynchronized states likewise occur in both statistical regimes.

These results demonstrate that, for the present Kerr-driven quantum van der Pol oscillator, photon statistics do not determine the onset of synchronization. The synchronization boundary originates from the competition between the Kerr-induced effective detuning and the squeezing-induced phase stabilization, both of which are encoded in the spectral response of the oscillator. By contrast, the Mandel parameter measures the fluctuations of the photon-number distribution and therefore probes a fundamentally different property of the quantum steady state. Although both observables are influenced by the same nonlinear dynamics, they evolve largely independently across the parameter space.

The absence of a one-to-one correspondence between synchronization and photon statistics highlights the fundamentally dynamical nature of quantum synchronization. While nonclassical photon statistics provide valuable information regarding the quantum fluctuations of the steady state, they cannot serve as a reliable indicator of phase locking. Instead, synchronization is more naturally characterized through phase-sensitive observables such as the power spectrum, the observed oscillation frequency, and the spectral linewidth. Photon statistics therefore provide complementary information that enriches the physical characterization of the steady state without determining the synchronization transition itself.

\section{Conclusion}
\label{sec:conclusion}

We have investigated the interplay between Kerr nonlinearity and parametric squeezing in a driven quantum van der Pol oscillator, combining semiclassical analysis with a full quantum treatment of the steady-state dynamics. Our results show that the Kerr interaction introduces an amplitude-dependent frequency shift that fundamentally modifies the synchronization properties of the oscillator. At the semiclassical level, this nonlinear frequency pulling drives a saddle-node bifurcation that transforms the phase-space structure from a bistable to a monostable regime. In the quantum regime, the same transition manifests itself through systematic shifts of the emission spectrum, linewidth broadening, and continuous deformations of the steady-state Wigner function, establishing a clear correspondence between the classical bifurcation picture and the quantum dynamics.

A central result of this work is the construction of quantitative synchronization phase diagrams over the two-dimensional parameter space spanned by the squeezing strength and the Kerr nonlinearity. From these diagrams, we extracted the critical synchronization boundary and demonstrated that, over a broad parameter regime, it follows a remarkably simple linear scaling law, where the required squeezing strength increases linearly with the Kerr nonlinearity. This proportionality specifies the minimum squeezing strength needed to compensate the Kerr-induced frequency shift and maintain phase locking, providing a practical calibration rule for controlling synchronization in nonlinear quantum oscillators.

To further characterize the quantum steady state, we analyzed the Mandel parameter throughout the synchronization phase diagram. Although the Kerr nonlinearity continuously modifies the photon-number statistics, we found that the synchronization boundary does not coincide with the crossover between super- and sub-Poissonian statistics. Instead, synchronized and unsynchronized steady states can each exhibit either type of photon statistics depending on the operating point. This demonstrates that photon-number statistics and synchronization describe fundamentally different aspects of the steady state: the former characterizes number fluctuations, whereas the latter is an intrinsically dynamical phenomenon governed by phase locking.

Taken together, the semiclassical fixed-point analysis, the Wigner-function evolution, the spectral characterization, and the photon-statistics analysis provide a unified physical picture of synchronization in the presence of strong Kerr nonlinearity. Rather than merely perturbing the synchronized state, the Kerr interaction reorganizes the synchronization landscape by introducing a controllable nonlinear detuning whose effects can be systematically compensated through parametric squeezing.

The present results establish a quantitative framework for engineering synchronization in nonlinear quantum oscillators and provide experimentally relevant design principles for platforms where Kerr nonlinearities naturally arise, including superconducting microwave circuits \cite{37,38}, integrated nonlinear photonic resonators \cite{39}, trapped-ion systems \cite{40}, and cavity optomechanical devices \cite{41}. More broadly, our work demonstrates how nonlinear frequency engineering can be exploited to manipulate synchronization in open quantum systems, offering a promising route toward robust phase control in future quantum technologies.

\bibliography{mybibb.bib}
\end{document}